# How does Module Tracking for Agrivoltaics Differ from Standard Photovoltaics? Performance & Technoeconomic Implications

Habeel Alam and Nauman Zafar Butt
Department of Electrical Engineering,
Lahore University of Management Sciences (LUMS)

*Abstract*— Spatial-temporal sharing of sunlight between solar modules and crops needs to be designed optimally in agrivoltaics (*AV*). For *AV* with fixed module tilts, the sunlight balance is governed through the spatial density and elevation of the modules which cannot be manipulated after the installation. For a flexible food-energy balancing across various seasons and crop rotations, modules with single or dual axis mobility can be best suitable. *AV* tracking must be geared towards ensuring a desired sunlight balance that may depend on many factors including the crop type, module array density, socio-economic factors, and local policies. Here, we explore single axis customized tracking (*CT*) for the mobile *AV* using a techno-economic model that incorporates design parameters including crop's shade sensitivity, module to land area ratio, and module types, as well as the economic parameters including soft and hardware costs for modules, feed-in-tariff, and crop income. *CT* is implemented through standard tracking that tracks the sun around noon hours and its orthogonal, $i.e.$, anti-tracking around sunrise/sunset. We evaluate the optimal *CT* schemes that can maximize economic performance while ensuring the desired food-energy yield thresholds. Economic feasibility for *AV* is evaluated in terms of the ratio ($ppr$) of the price for the module system customizations to the performance benefit due to the crop income. A case study for Punjab, Pakistan shows that *CT* schemes for moderate shade sensitive crops and typically dense *AV* module arrays can require 30 to 40% increase in the reference *FIT* to ensure the food-energy yield threshold of 80% relative to standalone food-energy farms for high and low value crops, respectively. *CT* schemes for a lower crop yield threshold of 70% require the corresponding increase in *FIT* to 10 to 20%, respectively. The proposed approach can be very effective for design and analysis of tracking schemes for *AV* systems.

*Index Terms*—techno-economic model, agrivoltaics, feed-in-tariff, customized tracking, economics, energy yield, food yield

## I. Introduction

The global pursuit of sustainable energy production has led to significant advancements in photovoltaic (*PV*) technology, making it a pivotal player in the transition to clean energy sources. Harnessing solar energy through *PV* systems is not only crucial for addressing the escalating energy demand but also mitigating climate change by reducing greenhouse gas emissions [1]. Rapid proliferation of ground-mounted photovoltaics (*GMPV*), while promising clean energy generation, has however ignited a pressing land-use conflict. Agricultural land, already under duress due to factors such as climate change impacts and urbanization, faces the additional challenge of accommodating expansive *PV* installations [2-5]. In this context, an innovative solution that reconciles energy production and agriculture, ensuring food security and sustainable energy generation, is of paramount importance.

Agrivoltaics (*AV*), emerges as a compelling solution to this conundrum [4] by enabling the coexistence of agriculture and solar energy production on the same land [6-8]. This integrated approach holds promise in alleviating land-use conflicts, optimizing resource utilization, and reducing greenhouse gas emissions[9]. The concept of *AV* was initially proposed by Goetzberger and Zastrov back in 1981 [10], but due to poor efficiency of photovoltaics at that time, this concept only became popular in last decade [11] initially explored in simulations in the start of last decade [8] and later implemented in many academic and commercial installations for different locations and crops across the globe [12-16].

Currently more than 3000 *AV* systems with cumulative capacity greater than $14 GW_p$ have been installed across the globe [12-14, 17]. Design parameters including modules' pitch, elevation, tilt angle have been explored [18-25]. The results indicate many synergies in the food-energy-water nexus including increased water use efficiency, higher crop yields for specific crops, improvement in microclimate, and a cooling effect on solar modules resulting in enhanced energy production. These benefits prompted governments in different countries like Germany (1982), Japan (2004), US (2008), China (2016), India (2018) and South Korea (2019) to develop and adopt policies supporting implementation of Agrivoltaics [26, 27].

Like commercial *PV* installations, agrivoltaics installations can be either fixed tilt or tracking technology. Although fixed tilt *AV* requires a lower capital cost as compared to the tracking systems, solar tracking for agrivoltaics can provide a higher flexibility to adjust the sunlight balance between crops and modules [28]. In the case of commercial *PV* installations, single or double axis solar trackers dynamically adjust the module orientation to perpetually face the sun to maximize energy capture. Tracking solutions for *AV*, however, requires models that can address the distinctive challenges and considerations to cater for a broad range of crop shade response and food-energy yield requirements. The standard solar tracking ($ST$) for $AV$ can result in a drastic reduction the crop yield which is contrary to the purpose of agrivoltaics. By using customized (also known as controlled or smart) tracking ($CT$), which utilizes both $ST$ and its orthogonal, *i.e.*, anti-tracking ($AT$) at different time intervals along the day, the crop and energy yield constraints are achievable through an optimized design of the tracking scheme.

Despite its importance, customized tracking for $AV$ is relatively less explored and reported. Valle et. al introduced the concept of customized tracking ($CT$) briefly and Elamri et. al explored the concept of control tracking (customized tracking) by developing a model which simulated the effects of fluctuating radiation and rain redistribution by the solar panels on crop growth, yield and water consumption [28, 29]. Their $CT$ scheme minimizes radiation interception before noon and after afternoon but shades the crops during the hot hours. Hussnain et al optimized the design of $AV$ systems, such as the spatial density, orientation, and tracking of the module arrays, according to the photosynthetic needs of different crops [30]. More recently, Willockx et. al explored the performance of a fixed vertical system and a dynamic single-axis tracker in Belgium with sugar beet cultivation [31] with theoretical modeling and field measurements over two growing seasons. The tracking system outperformed the fixed vertical system in both energy yield (+30%) and land use efficiency (+20%), mainly due to its ability to optimize the module position and shade levels for the crops based on time and location.

While the above-mentioned studies have validated the benefits of tracking for $AV$ in terms of land usage efficiency, increased crop yield for specific crops, and efficient water utilization, the economic and financial modeling which is crucial for policy makers and social acceptance is rarely reported for tracking $AV$ systems. Nevertheless, a few studies have been reported for fixed tilt $AV$ economic modeling and field experiments. Schindle et al reported a simple model based on price performance ratio and compared economic performance of winter wheat and potatoes in an $AV$ system with $GMPV$ [12]. The higher $LCOE$ for $AV$ was considered the price while the revenue from crops was considered as performance benefit. The study revealed that a high revenue from potatoes could offset the higher $LCOE$ of $AV$ and could make it profitable in comparison with $GMPV$ even with reduction in biomass yield of potatoes to ~87% with respect to full sun condition. Winter wheat, on the other hand, could not achieve economic feasibility due to lower profits from crops. Ryyan et al presents a numerical model by using a performance indicator based on economics, not land equivalent ratio (LER), to evaluate and optimize the $AV$ system with paddy rice, for six different locations across the globe [32]. It finds that $AV$ can provide 22-132 times higher profit than conventional rice farming while maintaining 80-90% of rice production.

A recent field study in Germany provides economic analysis of agrivoltaics in apple farming based on three pilot projects [33]. Using different calculation methods to assess the costs and benefits, the study finds that $AV$ can reduce the investment and operational costs of the apple farming system by 26% and 8%, respectively. It however can decrease the apple quality and revenues by 10% and 8%, respectively. The investigation in [24] delves into the economic performance of $AV$ relative to the roof top and $GMPV$ configurations. The study reveals that $GMPV$ systems exhibit a cost advantage of approximately 33% over $AV$ systems due to reduced expenditures but net present value ($NPV$) for $AV$ systems may ultimately yield a higher level of profitability by the end of project lifetime. In [34], [35], an economic framework (FEADPLUS) is presented to evaluated from the perspective of maintaining the profitability of farmer. The framework however misses the impact of land preservation cost on the profitability of solar investor with respect to $GMPV$.

The above-mentioned studies are although useful, their focus is limited and does not incorporate the combined effect of varying the module design, land costs, crop rotations, and $FIT$. In particular, the economic tradeoffs for various tracking options for $AV$ modules for different types of crops, soft and hardware costs have not been investigated. A holistic model is needed to explore the effect of shade sensitivities of different crops on tracking schemes and varying module configurations for $AV$ to meet the food and energy constraints. We have recently presented a technoeconomic model [36], which explores the aforementioned aspects for the for fixed $AV$ modules including $N/S$ faced and vertical bifacial $E/W$ faced configurations. In this paper, we extend the framework for tracking $AV$ systems and explore the design of efficient tracking schemes in terms of food-energy yield requirements and the economic performance. In addition, we evaluate the known economic parameters such as price and performance benefits in terms of system parameters including the hardware and soft costs, energy yields, land to module area ratio, and $FIT$.

Specifically, we develop a techno-economic model that addresses the following questions for the design and performance of tracking $AV$ in this paper: How a variety of shade response for crops influence the $CT$ schemes? (ii) What is the impact of energy and crop yield thresholds on the design of $CT$ schemes. (iii) What $CT$ schemes can be economically feasible relative to standard standalone food-energy systems while ensuring the desired food-energy thresholds, (iv) How the module array design in terms of land to module area ratio influence the techno-economics?, (v) What are the required Feed in tariffs for the mobile $AV$ systems with $CT$ for the crops having different market values (vi) How $CT$ schemes vary across various global locations for a given system design and food-energy thresholds.

The rest of the paper is arranged as follows: In section II, we report the methodology and mathematical modelling of this techno-economic framework highlighting its major assumptions and components. In Section III, we apply the framework to assess the economic feasibility of $AV$ for different tracking orientations ($ST$, $AT$ and $CT$) across two simulated crop rotations for Khanewal located in Southern Punjab, Pakistan. Section III presents results and discussion while discussing questions (i)—(vi) as mentioned in the preceding paragraph and. Finally, Section IV reports conclusion and limitations.

**Nomenclature**

| | | | |
|---|---|---|---|
| $A_L$ | Land area | $LV$ | Low value crops |
| $A_{LM}$ | Land to module area ratio | $m$ | meter |
| $A_M$ | Module area | $M_L$ | Module-to-land cost ratio |
| $AT$ | Anti-tracking | $n$ | Number of solar tracking hours |
| $AV$ | Agrivoltaics | $NPV$ | Net present value |
| $C_L$ | Soft costs | $p$ | Price |
| $c_L$ | Soft costs per unit land area | $p'$ | Normalized price |
| $C_M$ | Hardware costs | $PAR$ | Photosynthetically active radiation |
| $c_M$ | Hardware cost per unit module area | $pb$ | Performance benefit |
| $CT$ | Customized tracking | $pb'$ | Normalized performance benefit |
| $d$ | Depreciation rate | $P_C$ | Normalized crop profit |
| $FD$ | Full density ($A_{LM} = 2$) | $P_{C_{AV}}$ | Annual crop profit in $AV$ (\$/year) |
| $FIT$ | Feed in tariff | $P_{e_{AV}}$ | Annual energy profit from $AV$ |
| $GMPV$ | Ground mounted photovoltaics | $P_{e_{GMPV}}$ | Annual energy profit from $GMPV$ |
| $GW_P$ | Gigawatt-peak | $PPR$ | Price performance ratio |
| $h$ | Height | $PV$ | Photovoltaics |
| $ha$ | Hectare | $r$ | Discount rate |
| $HD$ | Half density ($A_{LM} = 4$) | $S$ | Shade sensitive crop |
| $HV$ | High value crops | $ST$ | Standard tracking |
| $\kappa_L$ | Normalized soft cost ratio | $T$ | Shade tolerant crop |
| $\kappa_M$ | Hardware cost ratio | $TD$ | One-third density ($A_{LM} = 6$) |
| $L$ | Shade loving crop | $Y_{Crop}$ | Biomass/crop yield |
| $LCOE$ | Levelized cost of electricity | $Y_{PV}$ | Energy yield |
| $LER$ | Land equivalent ratio | $YY$ | Energy yield per module area |
| | | $YY_T$ | Annual energy production |

II. MATHATICAL MODELLING

*A. Customized Tracking (CT) Model*

Agrivoltaics systems can be categorized into two types based on their configurations a) fixed tilt systems which include $N/S$ faced fixed tilt system and vertical tilt bifacial $E/W$ faced system b) tracking systems which incorporate trackers and use tracking strategies such as standard tracking ($ST$) and anti-tracking ($AT$). For **$N/S$ faced fixed tilt** orientation the modules are elevated at the height of 3-5 m and face $N/S$ direction while bifacial panels are installed vertically facing $E/W$ direction and generally elevated at the height of 1 m (crop height) for **$E/W$ faced vertical bifacial** orientation. The spatial light distribution over crops is more homogeneous than $N/S$ fixed tilt system but the energy generation is also lower[18]. In case of **Standard tracking ($ST$)** scheme, $PV$ modules track the sun prioritizing energy generation over food production. **Anti tracking ($AT$)** as the name suggests is opposite of $ST$ in such a way that in $AT$, module face is kept parallel to direct beam throughout the day prioritizing food production over energy generation. Fig. 1 shows conceptual design of the $N/S$, vertical bifacial $E/W$, $ST$ and $AT$ orientations for $AV$.

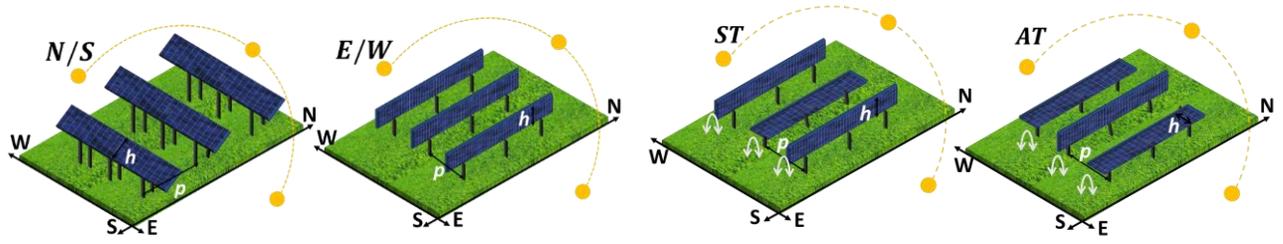

**Fig. 1.** Typical $N/S$ fixed tilt, $E/W$ vertical bifacial, Standard tracking ($ST$) and Anti tracking ($AT$) $AV$ systems with pitch ($p$) and height ($h$) labelled.

The $ST$ and $AT$ may not be the best techno-economic approach for agrivoltaics because while $ST$ maximizes the overall energy performance of $AV$ systems, agriculture production is decreased and may not be acceptable. $AT$ on the other hand, provides an agricultural production close to the full sun condition but significantly reduces the energy yield. Customized single axis solar tracking ($CT$) scheme is defined by multiplexing $ST$, which maximizes the energy, with anti-tracking ($AT$) which maximizes the agricultural yield. $CT$ incorporates both $ST$ and $AT$ such that $ST$ is implemented for $n$ hours with $n/2$ number of hours on each side of midday (noon) while $AT$ is implemented for the rest of the day as shown in Fig. 2.

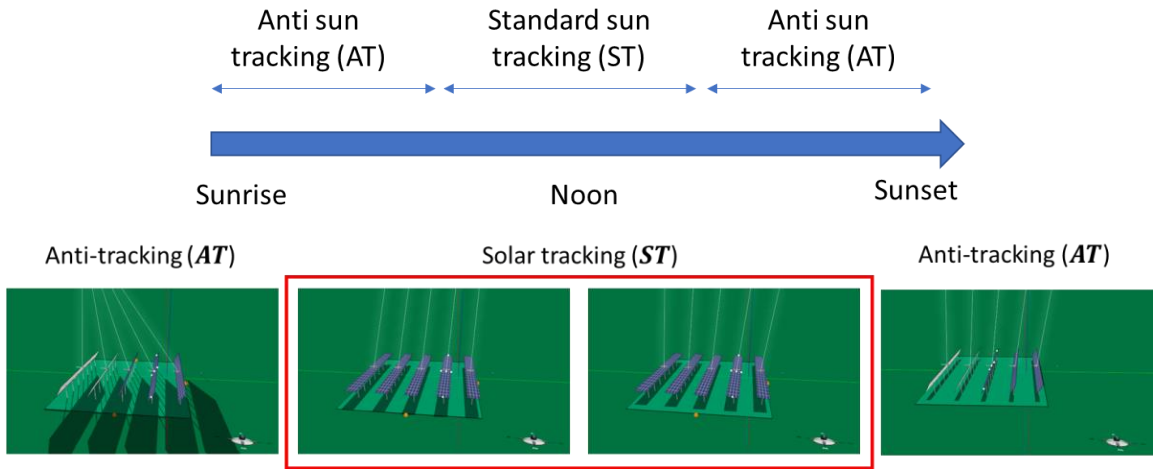

**Fig. 2** Customized tracking scheme illustration which utilizes solar tracking at noon while anti tracking for rest of the day to meet food and energy constraints.

*B. Energy and shading Model*

In our previous publications [19], [20], we explained the model for simulating energy generation within photovoltaic modules and the photosynthetically active radiation ($PAR$) available to crops beneath these modules. Here, we provide a concise overview of our methodology. Assuming relatively large arrays of modules and neglecting edge effects, we address shading patterns in two spatial dimensions, namely, perpendicular to the arrays' length and the height above ground. We employ a validated view factor model, established through field experiments [19], [20], [32], to compute sunlight interception by the modules, thereby determining the temporal $PV$ yield. This calculation encompasses contributions from direct sunlight, diffused light, and albedo (both direct and diffuse components). To ascertain the $PAR$ reaching the crops, we compute shading for direct and diffused light within 2-D vertical planes beneath the modules. Our simulations utilize typical meteorological data for Khanewal, Punjab, Pakistan (30.2864 °N, 71.9320 °E) [32], [36]. The model is used to compute energy yield ($Y_{PV}$) which is the ratio of energy yield per unit module area of a given $AV$ orientation to the energy yield per unit module area of GMPV. The model also evaluates the shading ratio which determines the light availability to the crops is the ratio of light available on the ground with modules installed to the light available at ground without the modules.

*C. Shade Sensitivities for Crop*

Crop yield reduction resulting from shading is quantified by assessing the decrease in the photosynthetically active radiation ($PAR$) received by the crop throughout the day. $Y_{Crop}$ is defined as the percentage of the biomass yield for a crop under shading to biomass yield of the same crop under no shading condition. $Y_{Crop}$ as a function of the $PAR$ availability for the crops relative to full sun condition has recently been analyzed in a meta-analysis with data from 58 studies [37]. Fig. 3 shows the response of $Y_{crop}$ to $PAR$ from the results reproduced from [37]. Crops having different shade sensitivities are classified as (i) shade sensitive ($S$) which are highly susceptible to shade, (ii) shade tolerant ($T$) which are moderately affected by shade, and (iii) shade loving ($L$) which are mildly affected by shade.

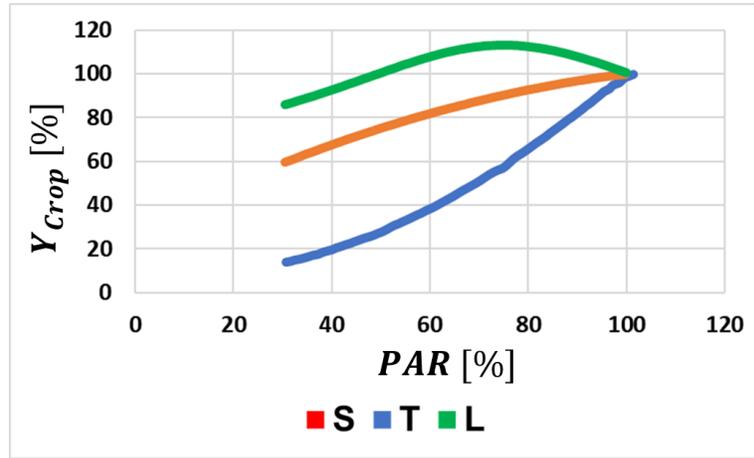

**Fig. 3.** Shade response for the crops of various shade sensitivities adapted from [37]

Land to module ratio ($A_{LM}$) is a parameter which depicts the module spatial density for agrivoltaics. For a given total area of modules, higher module density results in lower $A_{LM}$ and thus higher shading (lower $PAR$). Fig. 4 shows the impact of various crop sensitivities in form of bars over the range of land to module area ratio on four different $AV$ orientations a) $N/S$ fixed tilt, b) E/W vertical, c) $ST$ and d) $AT$ for Khanewal, Pakistan. $Y_{crop}$ for the crop types $S$, $T$, and $L$ is shown. For crops type $S$, $Y_{crop}$ is \mildly affected by land to module area ratio for all module configurations. For crop type $S$, $Y_{Crop}$ is heavily dependent on module configuration as well as on the land to module area ratio as both of the factors contribute to the shading ratio. $Y_{Crop}$ increases for shade sensitive crop with increase in $A_{LM}$ from full density ($FD, A_{LM} = 2$) to one-third density ($TD, A_{LM} = 6$). In terms of module configurations, $AT$ is best suited for shade sensitive crops, followed by E/W vertical and the fixed tilt $N/S$ orientations.

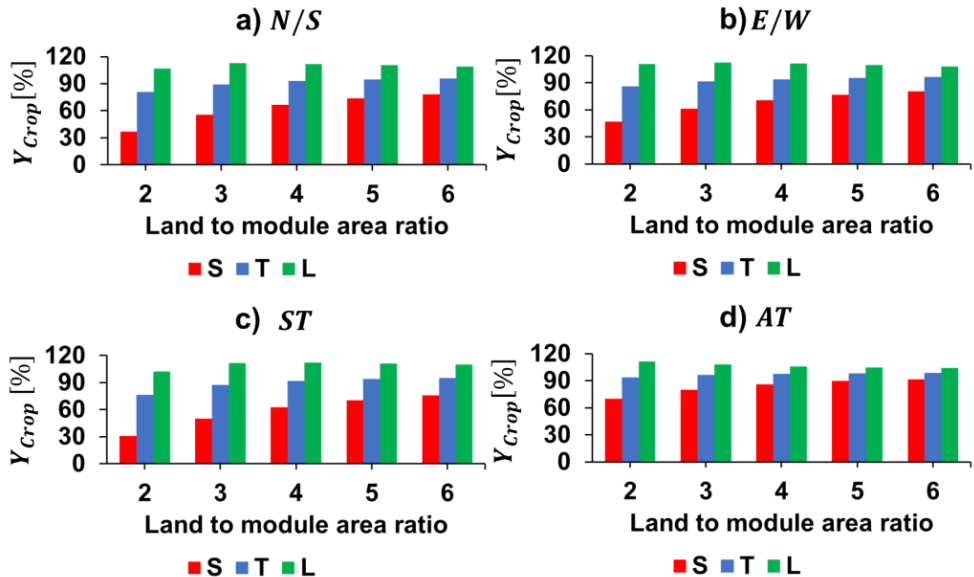

**Fig. 4.** Effect of different module orientations (a) $N/S$ fixed tilt, b) $E/W$ vertical bifacial, c) $ST$ d) $AT$) on $Y_{Crop}$ for different crop sensitivities for range of land to module area($A_{LM}$) ratio for Khanewal, Pakistan. Biomass yield ($Y_{Crop}$) increases with increase in $A_{LM}$ irrespective of crop sensitivity and orientations. $AT$ is best performing for lower $A_{LM}$ followed by vertical $E/W$, then $N/S$ fixed tilt. $ST$ is not recommended for shade sensitive crops at $FD$ and $HD$ for $AV$.

Fig 5. shows the annual values of $Y_{Crop}$ and $Y_{PV}$ for Rabi and Kharif seasons as a function of $ST$ hours along the day for three different crop shade sensitivities for three different land to module area ratio ($A_{LM} = 2(FD), 4(HD)\ and\ 6(TD)$). Both $Y_{PV}$ and $Y_{Crop}$ threshold criteria considered here is 80%. To support he $Y_{PV}$ criteria, $ST$ in a day must be 5 hours or greater as highlighted by the light green shaded region. When land to module area ratio ($A_{LM}$) is 2, Crop type $L$ cannot be supported with any $CT$ scheme while crop $T$ can be supported with $ST$ of 5-8 hours in a day. The crop $L$ can be supported with $ST$ of $5 - 12$ hours or more. For land to module area ratio of 4, crop types $L$ and $T$ can both be supported with $ST$ of 5 – 12 hours or more while crop $S$ cannot be supported. When land to module area ratio is 6, crop types $L$ and $T$ can both be supported with $ST$ of 5 – 12 hours or more while crop $S$ can be supported with $ST$ of 5-8 hours in a day. It should be noted that beyond a critical value of $ST$ hours in a day which is around 10 hours, $Y_{Crop}$ and $Y_{PV}$ tends to saturate. Below the $ST$ of 10 hours in a day, the variation in $Y_{PV}$ is significantly large as compared to that for $Y_{Crop}$.

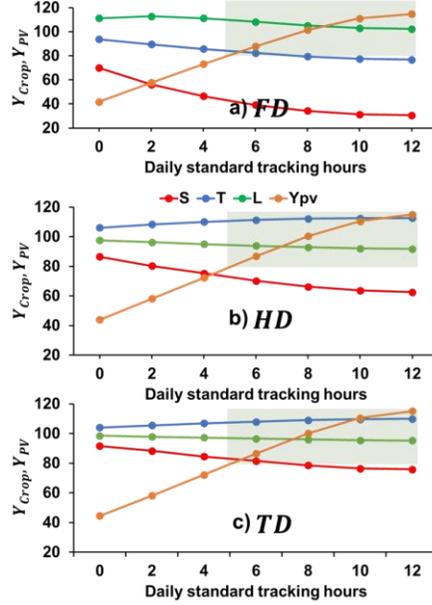

**Fig. 5.** Annual variation in $Y_{PV}$ and $Y_{Crop}$ for shade tolerant, shade sensitive and shade loving crops as function of daily standard tracking hours for full density ($FD, A_{LM} = 2$), half density ($HD, A_{LM} = 4$) and one-third density ($TD, A_{LM} = 6$). By increasing the land to module area ratio from 2 to 6, the $Y_{Crop}$ constraint of 80% for shade sensitive ($S$) crop can be achieved for $ST$ hours of 6-8 while the number of $ST$ hours for shade tolerant ($T$) and shade loving ($L$) crops also increase with increase in $A_{LM}$.

*D. Economic Model*

We use price-performance ratio ($ppr$) as a benchmark to evaluate the economic performance of $AV$. This model is based on our recent work [36] with extensions necessary for $ppr$ based analysis. The price and performance factors can widely vary according to the business scenario and the land/system ownerships. [12] describes five scenarios based on the several cooperation models between land users that include $PV$ operator, farmer, and the landowner. Although multiple business scenarios can exist in $AV$ between farmer, $PV$ investor, and the landowner, here we primarily focus on the case when the farming and $PV$ investments are owned by a single entity so that the maximizing of the overall profit is the main objective. The other scenarios where the $PV$ and the farming investments are shared between multiple owners, the model can be extended and applied according to the specific details of the business contract.

Typically, the hardware customization (i.e., the elevated mounting and stronger foundations) are the main contributors to the $AV$ price while the soft costs ($EPC$, taxes, and land lease, etc.) may have a relatively small difference as compared to the standard $GMPV$. While the hardware costs ($C_M$) are usually modulated by global economics, soft costs ($C_L$) depend more on the country specific policies and can further depend on the type of land and business models. With the bifurcation of the levelized cost of electricity ($LCOE$) into hardware and soft costs, can be re-written as [36, 38]:

$$LCOE = \frac{C_M + C_L}{YY_T \cdot \chi} = \frac{c_M A_M + c_L A_L}{YY_T \cdot \chi} = \frac{M_L + A_L/A_M}{YY \cdot \chi / c_L} \quad (1)$$

where $A_M$ and $A_L$ are the total module and land areas for $AV$, respectively, and $YY_T$ and $YY$ are the total energy and energy production per module area, respectively. $\chi \equiv \sum_{k=1}^{Y}(1-d)^k(1+r)^{-k}$, where $d$, and $r$ are rates for depreciation and discount rates, respectively. ($M_L = c_M/c_L$) is the hardware to soft cost ratio, where $c_M$ and $c_L$ are the hardware costs per unit module area and soft cost per unit land area, respectively. It is an important quantity which can influence the relative price for the $AV$ system. $M_L$ is typically close to 10 in US and vary between 5 – 35 worldwide [39] as shown in appendix (Fig. A3).

*A.1 Technoeconomic modeling without policy intervention*

Here we assume that there are no subsidies from government for $AV$. The case with feed-in-tariff ($FIT$) incentive will be discussed later. The business case for this scenario can be made if the dual food-energy profit from $AV$ exceeds or equates the individual profits, had the land was utilized for a single use, i.e., either food or energy. Since the energy investment and the net energy profits are usually much larger as compared to that for the agriculture on a given land area, the business case can be written in comparison to the standard $GMPV$ system:

$$P_{e,PV} - P_{e,AV} \leq P_{c,AV} \tag{2}$$

where $P_{e,AV}, P_{e,PV}$ are the annual energy profit from $AV$ and $GMPV$, respectively, and $P_{c,AV}$ denotes the $AV$ profit from crops in \$/year. The left- and right-hand sides of (2) represents the price and performance benefit, respectively, for the case of single entity owned food-energy $AV$ business with respect to a standard $GMPV$ system for a given capacity of the energy generation. The price ($p$) can further be decomposed into hardware and soft cost components using (1):

$$p = \left(\frac{M_L + \frac{A_L}{A_M}}{\frac{YY_T}{A_{M_{AV}}} \cdot \frac{\chi}{c_L}}\right)_{AV} - \left(\frac{M_L + \frac{A_L}{A_M}}{\frac{YY_T}{A_{M_{GMPV}}} \cdot \frac{\chi}{c_L}}\right)_{GMPV} * YY_T \tag{3}$$

where $YY_T$ is the total annual energy production which is taken to be the same for AV and GMPV.

After some simplifications, (3) can be written as [36]:

$$p = \left[\left(\frac{c_{M_{AV}}}{c_{M_{GMPV}}}\right) + \left(\epsilon \frac{A_{LM_{AV}}}{M_L} \frac{c_{L_{AV}}}{c_{L_{GMPV}}}\right) - \left(\frac{A_{LM_{GMPV}}}{M_L} + 1\right) Y_{PV}\right] * \frac{c_{M_{GMPV}} \cdot A_{M_{AV}}}{\chi} \tag{4}$$

where $Y_{PV}$ which is the ratio of annual energy generated per unit module area for $AV$ to that for standard fixed tilt $GMPV$ is also equal to the total module area ratio for $AV$ to that for $GMPV$ since both systems are assumed to generate the same total annual energy. $Y_{PV} = 1$ if the $AV$ system has the same module tilt and orientation as that for the reference $GMPV$. $Y_{PV}$ can be greater than 1 if modules with tracking are used for $AV$, and $Y_{PV} < 1$ for the vertically mounted bifacial modules facing East/West. The terms $A_{LM_{GMPV}}$ and $A_{LM_{AV}}$ are the land to module area ratios for $GMPV$ and $AV$, respectively. $A_{LM_{GMPV}} \approx 2$ for conventional $GMPV$ while $A_{LM_{AV}}$ is usually greater than 2 so that excessive shading could be avoided for the crops. As noted previously $A_{LM_{AV}} \approx 2$ and 4 are sometimes referred to full density and half density $AV$ systems in literature.

The 1st and 2nd terms in (4) represent the difference in hardware and soft cost for $AV$ relative to standard fixed tilt $GMPV$. The practical value of the 1st term, i.e., $\frac{c_{M_{AV}}}{c_{M_{GMPV}}} \equiv \kappa_M$ depends upon specific economic details for a given $AV$ system. For example, $\kappa_M$ reported for ~5m elevated mounting is about 1.38 in one of the studies done in Germany [12] . Since trackers typically could increase the module hardware premium cost by ~20% [40], $\kappa_M$ for elevated AV with tracking could be higher than that for the elevated fixed tilt AV systems. The 2nd term in (4) contains the soft costs ratio ($\frac{c_{L_{AV}}}{c_{L_{GMPV}}}$) for the $AV$ module systems to that for $GMPV$ which incorporates the difference in their land lease cost, engineering, procurement, and construction ($EPC$) costs, and labor costs. It has an inverse dependency on $M_L$ which implies that the relative economic impact of soft costs reduces if the hardware to soft cost ratio for the system is higher for a system. $\epsilon$ in the 2nd term is a fraction that signifies how the soft costs scale when the land area for $AV$ is increased. $\epsilon$ is typically less than 1 and can be related to increase in the electrical wiring, $EPC$ costs, and labor when the land area is increased for a given total capacity of modules [41]. The 3rd term in (4) incorporates the effect of $Y_{PV}$ and has inverse proportionality with $M_L$ which implies that the relative economic effect of varying the energy produced per unit area for $AV$ vs. $GMPV$ diminishes as $M_L$ is increased.

Since the hardware costs often play a dominant role in the economic feasibility of $AV$, it is insightful to normalize the price relative to the hardware cost of the standard $GMPV$. The normalized price ($p'$) is given by:

$$p' = \frac{p}{A_{M_{AV}} \cdot c_{M_{GMPV}}/\chi} = \left[\left(\frac{c_{M_{AV}}}{c_{M_{GMPV}}}\right) + \left(\epsilon \frac{A_{LM_{AV}}}{M_L} \frac{c_{L_{AV}}}{c_{L_{GMPV}}}\right) - \left(\frac{A_{LM_{GMPV}}}{M_L} + 1\right) Y_{PV}\right] \tag{5}$$

The 1st and 2nd terms represent the difference in hardware and soft cost for $AV$ relative to $GMPV$. The 3rd term represents the impact of relative energy generation per module area for $AVS$ as compared to that for $GMPV$. The three terms in right hand side of (5) can be written in shorthand as:

$$p' = (\kappa_M + \kappa_L - Y'_{PV}) \quad (6)$$

For the ideal limit, when the energy generation per module area is the same and there is no difference in soft and hardware costs for $AV$ vs. $GMPV$, $p'$ approaches to zero. For practical cases considered in this study, $p'$ is typically between 0.4 – 0.8.

The performance benefit can be written as:

$$pb = Y_{crop} \times P_{c_{fullsun}} \quad (7)$$

Where $P_{c_{fullsun}}$ is the crop yield under the full sun condition and $Y_{crop}$ is the percentage biomass yield for AV relative to full sun.

To compute $ppr$, we divide the performance benefit with the same normalization factor as we have used for the price. The normalized performance benefit ($pb'$) can be written as:

$$pb' = \frac{pb}{A_{M_{AV}} \cdot c_{M_{GMPV}}/\chi} = A_{LM_{AV}} \left( \frac{Y_{crop} \times P_{c_{fullsun}}/A_{L_{AV}}}{c_{M_{GMPV}}/\chi} \right) \quad (8)$$

where the ratio in the brackets represents the crop profit earned from a unit area of land divided by the hardware cost of installing the same unit area of $GMPV$ module. $pb'$ is typically smaller than $p'$ and can vary across a wide range for low vs. high value crop. $pb'$ can have a wide range ranging from the order of 0.1 for some of the high value crops, such as the horticulture crops, to the order 0.001 for the low value crops.

The price-performance ratio ($ppr$) is given as:

$$ppr = \frac{p}{pb} = \frac{p'}{pb'} \quad (9)$$

Since the $pb$ can be much smaller than $p$ for many of the practical scenarios, it can be challenging to attain $ppr \leq 1$. This then necessitates some policy interventions to facilitate an economic viability for AV as discussed in the next section:

*A.2 Technoeconomic modeling with policy intervention*

When government incentives such as feed in tariff ($FIT$) are available, their economic impact be included in the performance term.

$$pb' = \left( \frac{pb/A_{L_{AV}}}{c_{M_{GMPV}}/\chi} \right) + \Delta FIT \left( \frac{YY_T \cdot \chi}{A_{M_{AV}} c_{M_{GMPV}}} \right) \quad (10)$$

where $\Delta FIT$ is the difference in $FIT$ for $AV$ and $GMPV$ and is assumed to be a positive number. For a given $AV$ system, a threshold $\Delta FIT$ can be computed to enhance $pb'$ so that $ppr'$ becomes close to one.

$\Delta FIT$ can be used as a tool by the policy makers to support agricultural land preservation through $AV$. Moreover, $\Delta FIT$ can be made crop-specific if cultivation of some selected crops needs to be promoted at a given location.

III. RESULTS AND DISCUSSIONS

The modeling framework is applied to a case study for two conceptual $AV$ farms: a) high value, and b) low value farms represent crop rotations that yield high and low annual profit, respectively for Khanewal (30.2864° N, 71.9320° E), Punjab, Pakistan. Each farm is studied under various CT schemes and compared to reference fixed tilt south faced $GMPV$. The cropping cycle and reported crop yield/revenues for Khanewal are taken into consideration while simulating the low value and high value farms. Crop rotation for the high value farm comprises of tomato, cauliflower, and garlic over the year, while for the low value farm, it consists of wheat and cotton as shown in Table I in appendix. These crops can be classified under shade tolerant crops based in their biomass yield ($Y_{Crop}$) as shown in Fig. A2 in appendix.

## A. CT for various crop types and seasons

$CT$ schemes can be optimized for a given crop type and season by adjusting the number of daily $ST$ hours centered around noon while doing $AT$ during rest of the day. Fig 6. shows the monthly values of $Y_{Crop}$ and $Y_{PV}$ for Rabi and Kharif seasons as a function of $ST$ hours along the day for three different crop shade sensitivities for land to module area ratio ($A_{LM}$). The Rabi season in Pakistan is from Nov-Apr while Kharif season is from May-Oct. Rabi crops for shade sensitive, shade tolerant and shade loving categories are shown Fig. 6 (a, c, and e), respectively, while Kharif crops for the same categories are shown in Fig. 6. (b, d, and f), respectively. The $Y_{PV}$ tends to saturate as the $ST$ hours go beyond 10 hours while the saturation of $Y_{Crop}$ curve is dependent on crop's shade sensitivity. To illustrate the feasible design space for the $ST$ hours in a day, we assume a case where the thresholds for $Y_{Crop}$ and $Y_{PV}$ of 80% need to be satisfied. The yellow and green shaded regions in Fig. 6 respectively represent the daily allowed $ST$ hours where $Y_{PV}$ and $Y_{Crop}$ thresholds are met across all the months in the season. An overlap between the two shaded regions corresponds to the tracking design for the daily $ST$ hours that could meet the energy and food constraints. It can be observed that $Y_{PV}$ threshold is met across all months with $ST > 7$ hours for both Rabi and Kharif, respectively. The $Y_{Crop}$ threshold, however, has a strong dependence on the shade sensitivity of the crop. For crop type $S$ (Fig. 6 a-b), the crop threshold is not met even with $AT$ (i.e., $ST = 0$) for the whole day for both the seasons. For crop type $T$ (Fig. 6 c-d), the $Y_{Crop}$ threshold is met for $ST \leq 7$ hours for both Rabi and Kharif. For these crops, the feasible tracking scheme is only when $ST$ in a day is around 7 hours where the required food and energy thresholds are barely met simultaneously across all months of the season. Finally, for crop type $L$ (Fig. 6 e-f), the food and energy thresholds are conveniently met irrespective of $ST$ hours and there is a complete overlap for all values of $Y_{PV}$ and $Y_{Crop}$ since the crop yield remains above 80% even when $ST$ hours are increased to 12. It should be noted that both $Y_{PV}$ and $Y_{Crop}$ show monthly variations across all types of crops and seasons. This is due to the natural variations in the sun's trajectory across months that change the shading ratio for the crop and solar energy generation. Although, the shaded regions in Fig. 6 are drawn with an assumption that the daily $ST$ hours are not designed to be changed across various months in each season, this is not an essential requirement in practical situations and is assumed here for simplicity. A monthly adjustment in $ST$ hours across the season can indeed be implemented to better facilitate the food-energy thresholds.

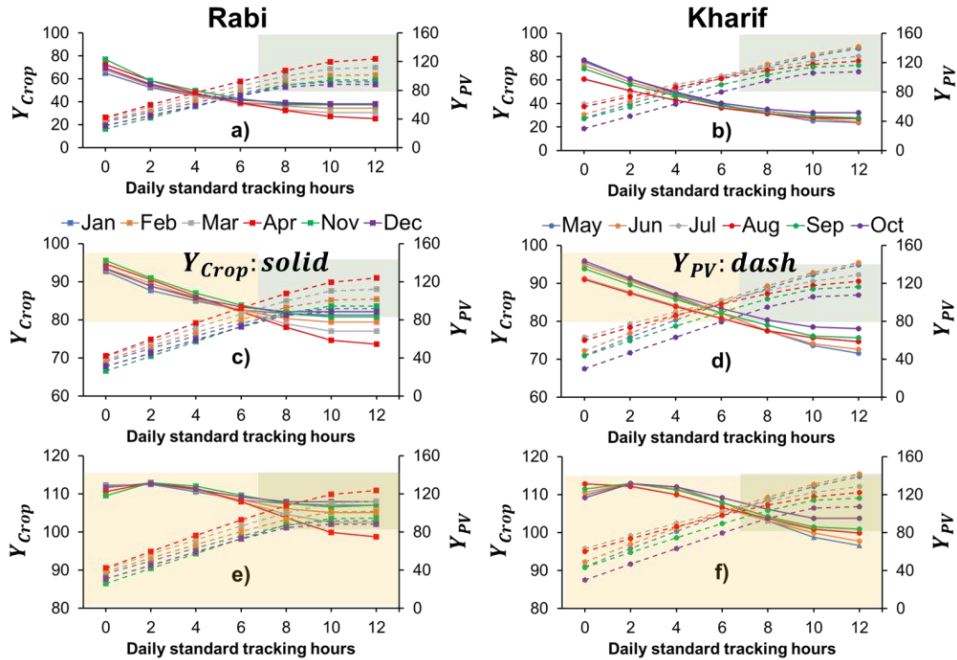

**Fig. 6.** Monthly $Y_{Crop}$ and $Y_{PV}$ are shown for Rabi and Kharif seasons as a function of $ST$ hours along the day for three different crop shade sensitivities at land to module area ratio of 2. The yellow and green boxes show 80% constraints for $Y_{Crop}$ and $Y_{PV}$, respectively. For the most shade sensitive crops, the constraints for the yield cannot be met at any value for $ST$ hours although the shade tolerant crop comes closer to the constraints for both seasons. The constraints are conveniently met for the shade loving crop for $ST$ hours of 8 or above.

## B. Impact of land to module area ratio on CT

The tracking scheme that can meet the thresholds for both crop and energy depends on the crop shade sensitivities as described in the previous section. At the system design stage, the land to module area ratio can be optimized by varying row-to-row spacing for the module arrays (assuming the land area for the system is adjustable) to allow for a lower shading ratio and a broader range of crops in the system. Fig. 7 shows how an increase of land to module area ratio from 2 to 6 can make $CT$ scheme viable for the crop type $S$ in both Rabi and Kharif seasons. The thresholds for $Y_{PV}$ and $Y_{Crop}$ are taken as 80% and 70%, respectively. For land to module area ratio of 2 (full density $FD$) and 4 (half density $HD$), there is hardly any $CT$ solution available to support the given food-energy thresholds except for the Rabi season where $ST = 7$ hours in a day can barely meet the thresholds with land to module area ratio of 4. On the other hand, when the land to module ratio is increased to 6 (one-third density $TD$), the food-energy thresholds are conveniently met for a broader range ($> 6$) of daily $ST$ hours across both seasons.

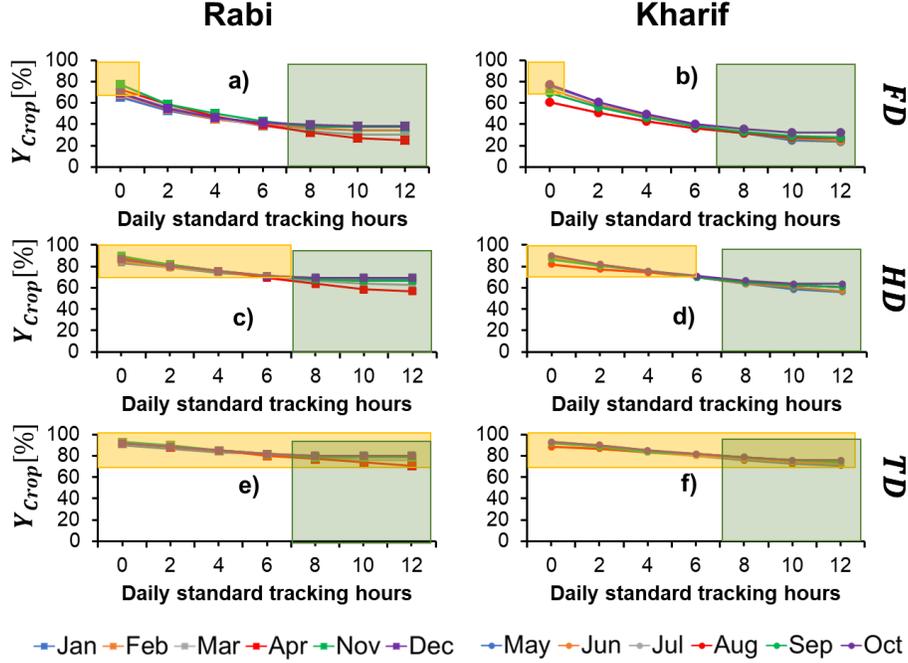

**Fig. 7.** Monthly values of $Y_{Crop}$ and $Y_{PV}$ are shown for Rabi and Kharif seasons as a function of $ST$ hours along the day for the most shade sensitive crop at land to module area density of 2 (Fig. 7a, b), 4 (Fig. 7c, d), and 6 (Fig. 7e, f). The yellow and green boxes show 70% and 80% constraints for $Y_{Crop}$ and $Y_{PV}$, respectively. The energy constraints are met for $ST$ hours=7 hours in a day as shown in Fig. 6. In a, b, and d, the energy and food thresholds are not simultaneously met at any value of $ST$ hours in a day. For c, the food-energy threshold are barely met at $ST = 7$ hours in a day. For Fig. 7e, f, the food-energy thresholds are met for all values of $ST$ hours greater than 7.

The above results highlight that crop and energy yield thresholds could be met either by selecting crops with an appropriate shade sensitivity for a given module configuration or by increasing the land to module area ratio at the design stage to allow for a broader crop types. If an AV system is already installed, then land to module area ratio is fixed and it cannot be altered. In this case, we can only customize the tracking for selected crops to meet the food-energy thresholds.

### C. Techno-economic modeling for the tracking AV

Till now, we have considered $CT$ from the perspective of fulfilling the food-energy thresholds for crops with different shade sensitivities and systems with varying land to module area ratio. In practice, however, the economic aspects could often play a decisive role in determining the $CT$ scheme. In this section, we will explore the economic performance of mobile $AV$ systems with various $CT$ schemes relative to the standard $GMPV$ system. System parameters including land to module ratio and daily $ST$ hours in a day are explored along with the economic parameters including crop profit and $FIT$ to quantify their effect on the economics. Only the crops having moderate shade sensitivities are considered in this section to keep the focus on the economic analysis. The approach is however applicable to any shade sensitivity for the crops. In the following sub-sections, we will first apply economic model on the standard $ST$ and $AT$ in comparison with the south faced fixed tilt $AV$ systems. We will then explore $CT$ schemes that can maximize the economic performance while ensuring food-energy yield thresholds.

### C.1. Effect of land to module area ratio

Fig 8 shows the how various economic paramters that define the price and perfromance benefit (eqs. (5) and (8)) depend on the land to module area ratio. Fig 8a shows that the hardware cost ratio remains constant as a function of land to module area ratio for both mobile and standard $AV$ systems. Hardware cost is higher for the mobile modules as compared to the fixed tilt orientation as expcted [41]. Fig 8b shows the effect of module to land area ratio on the 2$^{nd}$ term ($\kappa_L$) in the price equation (eq. (5)) that contains the effect of soft cost. $\kappa_L$ increases linearly with increase in land to module area ratio with the slope that depends on the scaling factor $\epsilon$ and $c_{L_{AV}}/c_{L_{GMPV}}$ (inset) as shown in Fig. 8b. Fig 8c shows the $pb'$ for low and high value crops which both increase linearly with land to module area ratio as more crops can be grown with increasing land. Moreover, the shading ratio for the crops reduces as the land to module area ratio is increased. The inset figure shows the zoomed plot for $pb'$ for the low value crops. Note that $pb'$ is much higher for high value crops in comparison with low value crops. For the lowest land to module area ratio, anti tracking shows a higher $pb'$ while at higher land to module area ratios, $pb'$ for all module configuration converge. This is due to the fact that significantly higher quantity of sunlight is available for crops with $AT$ as compared to $ST$ and fixed tilt system which results in higher $Y_{Crop}$. At higher land to module area ratios, the shading ratio becomes significantly lower for $ST$ and fixed tilt sysetms as well, and thus the anti tracking is not as beneficial as compared to other module configurations. Fig 8d shows the $Y_{PV}$ for different module schemes which remains constant irrespective of land to module area ratio. This is because energy yield per unit module area does not change with varying the land area unless there is mutual shading between the modules. For the range of module to land areas we have considered, mutual shading between modules is not significant. Figure 8d shows that $ST$ scheme generates the highets yield followed by $N/S$ faced fixed tilt modules while $AT$ scheme has the worst energy perfromance as most of the light is delivered to the crops.

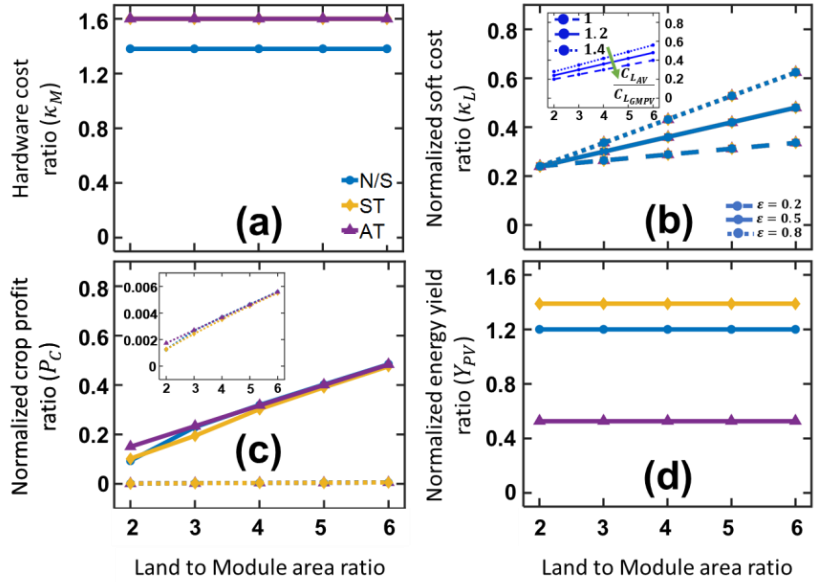

**Fig. 8.** Effect of land module area ratio ($A_{LM}$) on various economic parameters a) Hardware cost ratio ($\kappa_M$), b) Normalized soft cost ratio, ($\kappa_L$) c) Normalized crop profit ratio ($P_C$) and d) Normalized energy yield ratio ($Y_{PV}$) for $N/S$, $ST$ and $AT$ orientations for $AV$ for $M_L = 10$. $\kappa_M$ and $Y_{PV}$ remains constant irrespective of $A_{LM}$ for each orientation while $\kappa_L$ and $P_C$ shows increasing trend with increase in $A_{LM}$ for all the orientations. The insets of Fig 8b and c shows the impact $A_{LM}$ on land price ratio ($C_{L_{AV}}/C_{L_{GMPV}}$) and zoomed in $P_C$ for low value crops respectively.

Fig. 9 shows the price, performance and ppr for the $ST$, $AT$, and $N/S$ faced modules as a function of land to module area ratio considering the high value crops. Price and performance both increase linearly with increase in land to module area ratio of 3 and higher although the relative increase in the performance exceeds that for the price. This results in decrease in the $ppr$ as shown if Fig. 9c. As $ppr \leq 1$ is desired for economic feasibility, higher land to module area ratio tends to achieve this because of the increasing trend in the performance. Around the land to module area ratio of 6, ppr decrease tends to saturate while the economic feasibility, i.e. $ppr < 1$ is still not achieved. Compared to $ST$ and $N/S$ fixed tilt systems, $AT$ has the significantly higher ppr because of its lowest contribution in the energy yield and a high initial hardware cost.

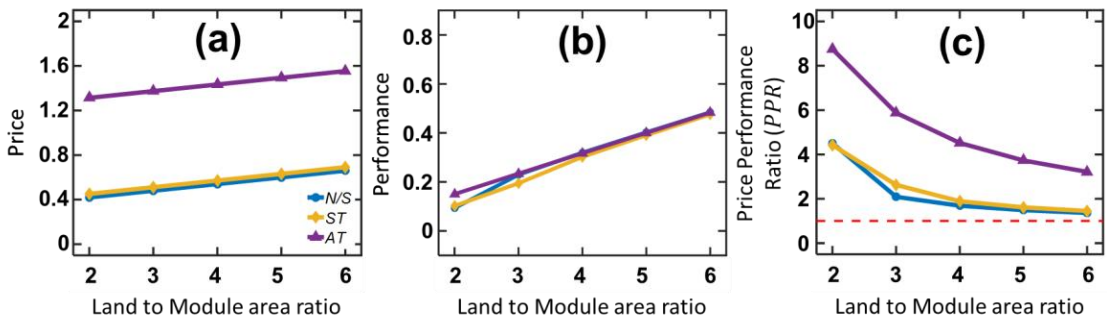

**Fig. 9.** Effect of land module area ratio ($A_{LM}$) on a) price, b) performance, and c) price performance ratio ($ppr$) for $N/S$, $ST$ and $AT$ orientations for $AV$ for high value crop (HV) at $M_L = 10$. Price and performance increase linearly with increase in $A_{LM}$ while $PPR$ decrease with increase in $A_{LM}$ for the three orientations explored. The economic feasibility ($ppr \leq 1$) is not achieved for any orientation.

*C.2. Effect of module soft to hardware cost ratio ($M_L$)*

Module hardware to soft cost ratio can have important implications for the economic feasibility of $AV$. Fig. 10 shows the effect of $M_L$ on price, performance and ppr for the $ST$ scheme and high value crop rotation. Lower $M_L$ therefore implies a higher soft cost and vice versa. Fig. 10a highlights that increasing $M_L$ lowers the slope of the price as a function of land to module area ratio. Higher $M_L$ results in decrease in $ppr$ and improves the economic viability of the standard tracking at higher land to module area ratios. These results highlight that when module to soft cost ratio is higher, increasing the land area (which mostly affects the soft costs) has a relatively mild impact on price. In contrast, when the module to land ratio is lower, increasing the land area (i.e., higher soft costs) has a stronger impact on price. Fig. 10c shows that with a higher $M_L$ of 30, $ppr$ can almost reach to its desired range of $\leq 1$ at land to module area ratio of 6.

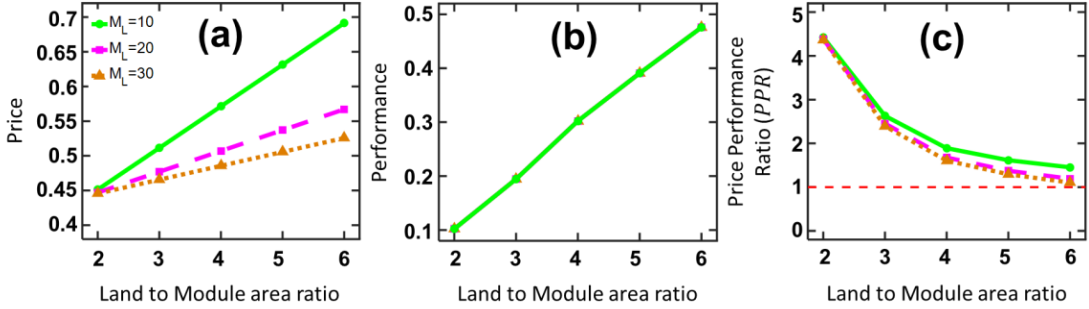

**Fig. 10.** Effect of module hardware to soft cost ratio ($M_L$) on a) price, b) performance, and c) price performance ratio ($PPR$) for $N/S$, $ST$ and $AT$ orientations of $AV$ for high value crop ($HV$) for land module area ratio ($A_{LM}$=2-6). Higher $M_L$ (lower land related costs) increases the slope of the price curve and thus decreases $ppr$, while performance remains unaffected with change in $M_L$. The economic feasibility ($ppr \leq 1$) is not achieved for smaller $M_L$ while $M_L = 30$ enables $ppr \sim 1$ for $A_{LM} = 6$.

*C.3. Effect of crop's market value*

Fig. 11 shows the effect of crop's value on the performance benefit and ppr as a function of land to module area ratio using high and low value crops. For low value crops, the performance benefit is significantly low (inset of Fig 11b) and economic feasibility is not achieved for any land to module area ratio. It should be noted that the curves of ppr for $N/S$ faced modules and $ST$ tend to saturate at higher $A_{LM}$ for both low and high value crops. For high value crops the economic feasibility is still not fully achieved for $ST$ and $N/S$ faced modules at module to land area ratio of ($A_{LM}$) 6 although the ppr comes close to 1.

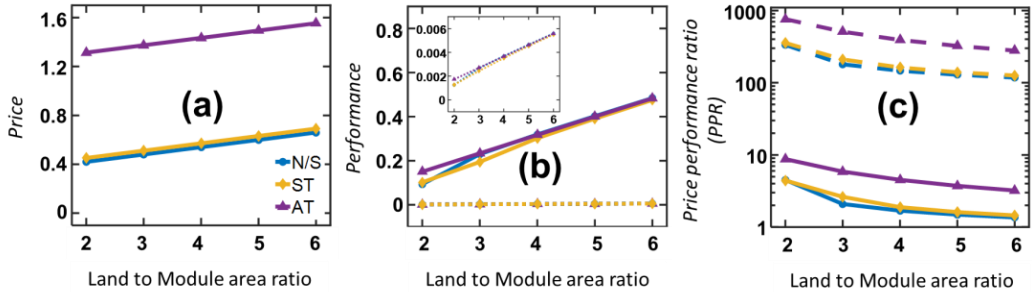

**Fig. 11.** Effect of crop's market value ($HV$ and $LV$) on: a) price, b) performance, and c) price performance ratio ($ppr$) for $N/S$, $ST$ and $AT$ module schemes for a range of land module area ratio ($A_{LM}$=2-6) at $M_L = 10$. The crop's profit impacts performance which increases linearly as $A_{LM}$ is increased which correspondingly decrease the $ppr$. Inset of Fig. 11b shows the zoom in performance for LV crop. A substantially smaller performance results in an extremely high $ppr$ for LV crops. The economic feasibility ($ppr \leq 1$) is not achieved for all types of module schemes although $ppr$ decreases significantly at higher $A_{LM}$ for high value crops.

*C.4. Effect of FIT*

Since economic feasibility is often not achieved even for high value crops, policy intervention in terms of subsidy, feed-in tariff, loans might be required to make $AV$ economically attractive to investors and farmers. The effect of $\Delta FIT$ on performance and ppr is incorporated in 10. Fig. 12 shows the effect of $\Delta FIT$ on ppr for high value crop and $M_L = 10$ varying the land to module area ratio from 2 to 6. $\Delta FIT$ impacts the performance curves, which shift upwards, while the ppr curves shift downwards with increase in $\Delta FIT$. $AV$ system for $N/S$ faced fixed tilt modules and $ST$ become economically feasible for $\Delta FIT = 10\%$ when their $ppr$ falls below 1. $AT$, on the other hand, will require a high value of $\Delta FIT$, (even greater than 30%) to become economically feasible. In case of N/S faced tilt modules and $ST$, $AV$ remains economically viable for all values of land to module area ratio for $\Delta FIT > 10\%$.

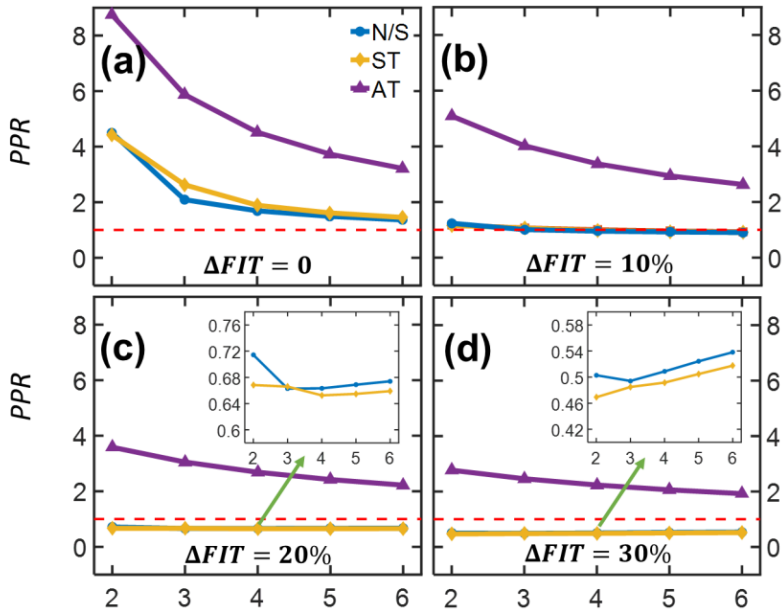

**Fig. 12**. Effect of $\Delta FIT$ on ppr for N/S, ST and AT for HV crop. Except AT and p/h=2, all the orientations become economically viable for $\Delta FIT \geq 10\%$. The insets in Fig. 12c and d shows zoomed in curves of N/S and ST which are already economically viable. The insets show an increasing trend with $A_{LM}$ due to higher profits from energy. AT might need higher $\Delta FIT > 30\%$ to become economically feasible due to poor energy yield.

*C.5. Economic impact of customized tracking*

As the $CT$ uses a combination of both $AT$ and $ST$ along the day, it can be explored to find the economic feasibility while either $ST$ or $AT$ fails to simultaneously meet the thresholds for both food and energy yield. Fig. 13 shows variation in price, performance and ppr for $CT$ schemes with respect to $ST$ hours in a day. The figure is drawn for land to module area ratio of 3 and $M_L = 10$ for high value ($HV$) and low value ($LV$) crops. A comparison for $\Delta FIT = 0$ and $\Delta FIT > 0$ cases is performed for both LV and HV crops which depicts similar trends. LV however requires a higher $\Delta FIT$ as that for HV to become economically feasible. For LV crop, $\Delta FIT = 30\%$ is required for economic viability for the daily ST hours$\geq 8$, while for HV crop, $\Delta FIT = 10\%$ enables economic viability with daily ST of 10 hours or higher. Since $ppr$ is the ratio of price and performance, the intersection of price and performance curves in Fig. 13c and 13d highlight the required $ST$ hours to obtain $ppr \leq 1$ hence making the AV economically viable.

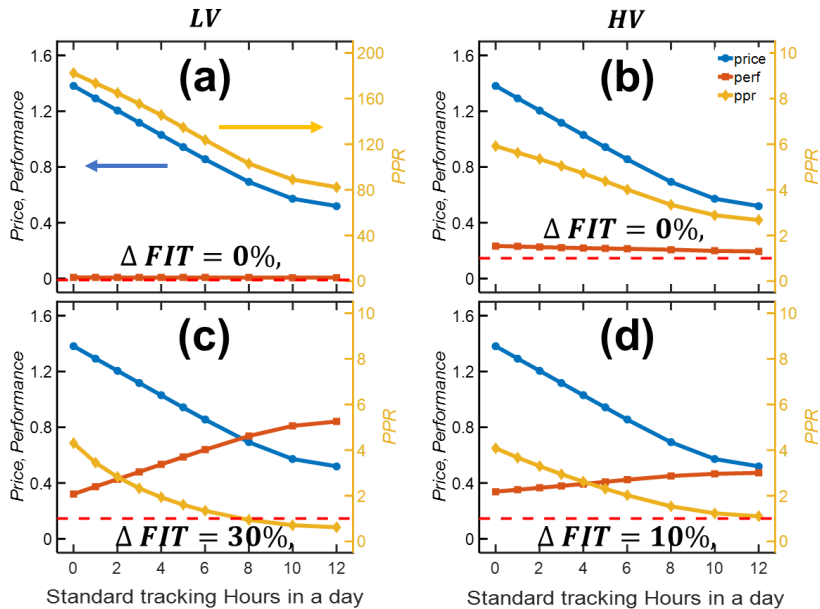

**Fig. 13.** Variation of price, performance and ppr with standard tracking hours in a day for p/h=3 and $M_L = 10$. Price, performance and ppr shows decreasing trend for $\Delta FIT = 0$ for both LV and HV crops. The performance shows increasing trend in case of $\Delta FIT = 30\%$ and $10\%$ for LV and HV crops respectively. The $ppr$ becomes economically viable when price and performance curves intersect eac-h other and performance becomes greater than price.

Fig. 14 shows the $\Delta FIT_{TH}$ (defined as the bare minimum $\Delta FIT$ that is required for $ppr \leq 1$) for $LV$ and $HV$ crops as a function of standard tracking hours in a day. Since the performance benefit for $LV$ is significantly low (implying high ppr) a greater $\Delta FIT_{TH}$ is required for it in comparison to $HV$ crops. As $ST$ hours increase from 0 (that corresponds to $AT$) to 12 (that corresponds to standard $ST$), $\Delta FIT_{TH}$ requirement for both $LV$ and $HV$ decreases. This is mainly because of energy yield and thus the energy profits becoming higher with increase in $ST$ hours. As discussed in previous section, however, increasing $ST$ hours for improving the economics must be limited to the constraints imposed by food-energy thresholds for the given $AV$ system. Higher energy profit might tempt the $AV$ investor to maximize the $ST$ hours which may decrease the crop yield drastically. This can be regulated by policy to curtail the $ST$ hours in a day to safeguard the food-energy thresholds. The $ST$ hours in a day should therefore be a crop and threshold dependent parameter, and it should be selected carefully in such a manner that both energy and crop thresholds are met in most economically beneficial way.

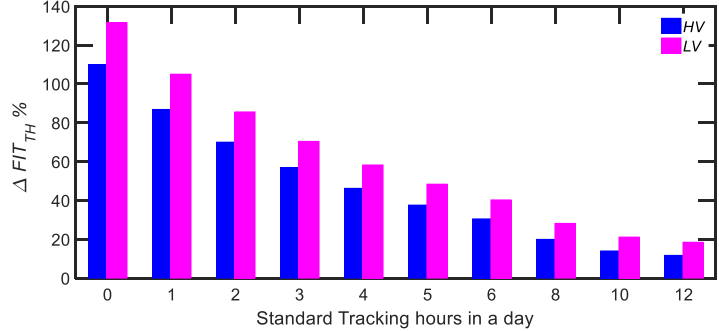

**Fig, 14.** Variation in $\Delta FIT_{TH}$ for HV and LV for $p/h = 3$ and $M_L = 10$ crops wrt standard tracking hours in a day. The $\Delta FIT_{TH}$ requirement decreases with increase in $ST$ hours.

*D. Limitations and future extensions*

While the criteria outlined in equations (9) and (10) provide accurate assessments, our model assumes the economic viability based on the premise that the same entity owns both the energy and food production. This assumption holds true in cases where a farmer is also the solar investor or vice versa, but it may not always be applicable. When the solar investor and the farmer are different entities, the profits generated from energy and crop yields, as well as the associated land costs, must be distributed according to their business arrangement. In such scenarios, government policy interventions become significantly more critical and can exert substantial influence on the technical and economic parameters. We are planning to extend our model to address these diverse scenarios as part of our future research.

## IV. CONCLUSIONS

In this paper, we have explored customized tracking ($CT$) for $AV$ through a techno-economic model. The $CT$ multiplexes the standard sun tracking (ST) with its orthogonal, *i.e.*, anti-tracking (AT) such that the ST covers noon hours and AT is done towards morning/evening. Economic feasibility is modeled using the price and performance benefit framework where price corresponds to the module system customizations required for $AV$ while the performance benefit is the crop income. The model computes the price separately for the soft and hardware components of $AV$ and incorporates for any difference in the energy produced per unit module area relative to the standard $GMPV$ configuration. Using the model, we explore the effect of crop's shade sensitivity, module type and areal density, and economic factors including the crop income and required feed-in-tariff. We show how the duration of ST hours in a day can be optimized to meet the threshold yield requirements for food and energy yield and to maximize the economic benefit. A case study for Lahore, Pakistan based on the model is presented with the following key conclusions:

- Combined food-energy yield requirements for the shade loving crops can be supported with the standard sun tracking across the whole day for land to module area ratio of 2 (full array density) or greater (reduced array density).
- Combined food-energy yield requirements for the crops with moderate shade sensitivity are barely met with standard tracking of 6 hours in a day with full array density. At reduced array density (land to module area ratio of 4 or more) standard tracking across the whole day can provide the required food-energy thresholds.
- Combined food-energy yield requirements for the crops with high shade sensitivity cannot be supported with full density arrays except with anti-tracking across the whole day. Half density arrays can barely support the food-energy requirements in some months for these crops with maximum standard tracking of 6 hours in a day. For further reduced density (land to module area ratio of 6) standard tracking across the whole day can support the food-energy thresholds for these crops.

- For high value crops, economic feasibility is not met without $FIT$ incentive even at reduced module array densities. A 10% increase in the reference $FIT$ for $GMPV$ can meet the economic threshold for high value crops with a slight reduction in the standard module array density.
- For low value crops, ~30% incentive in $FIT$ is required to meet the economic threshold for high value crops with a slight reduction in the standard module array density.
- The requirement of higher $FIT$ increases when the standard tracking hours in a day are reduced. The standard tracking hours are however required to be reduced when shade sensitivity of the crop demands smaller shading ratio to meet the crop yield threshold. The reduced standard tracking hours should nevertheless be compensated by a higher $FIT$ incentive for $AV$'s economic feasibility. In our case study for high value crops, ST of 12 hours and 6 hours in a day requires $FIT$ incentive of 10% and 30%, respectively. The respective $FIT$ incentives for the case of low value crops are 20% and 40%.

In summary, we show that techno-economic feasibility and design of module tracking for $AV$ can be customized using the presented model. Although the tracking infrastructure requires a high capital cost, it offers great flexibility to address the requirements of food-energy threshold yields. The economic performance is typically higher when standard tracking is done for most part of the day due to relatively high energy profits. This should however not be an acceptable solution when crops sunlight requirement is not met due to high shading. A model-based optimization for the standard tracking hours for the desired crop need is therefore a valuable solution.

## V. APPENDIX

### A. *Variation of CT across various global locations*

Since the daily and seasonal trajectory of sun varies with reference to the global coordinates, optimal $CT$ schemes can vary for different global locations. Fig A1 illustrates this behavior for four different locations (Khanewal, Heggelbach, Arizona and Sydney) where the latter is in the southern hemisphere. For each location, the maximum allowed $ST$ hours are evaluated using the approach described in Fig. A1 for the shade tolerant crop. The energy and crop yield thresholds set to 80% and land to module area ratio is 2. For the winter months (Nov-Feb) in the northern hemisphere, $ST \geq 12$ hours is possible for Lahore, Heggelbach and Arizona, while $ST$ for Sydney is limited to ~7 hours where it is summer as shown in Fig. A1(a). During the winter months (April-Aug) for the southern hemisphere, $ST \geq 12$ is possible. For months of Mar-Oct and Sep-Mar, locations in the northern and southern hemisphere show a slight variation in the maximum allowed daily $ST$ hours as shown in Fig. A1 (a).

Fig. A1 (b) shows a comparison of monthly $Y_{PV}$ for the four global locations when the $ST$ hours are customized as shown in Fig. A1 (a). The energy yield is highest in the month of June for Lahore for locations in northern hemisphere while for Sydney it is highest in December. These trends highlight that when $CT$ schemes are optimized across different global locations, the resultant food-energy yield may be slightly different across these locations. This insight may be useful when comparing the data from $AV$ system spread across different locations in the world.

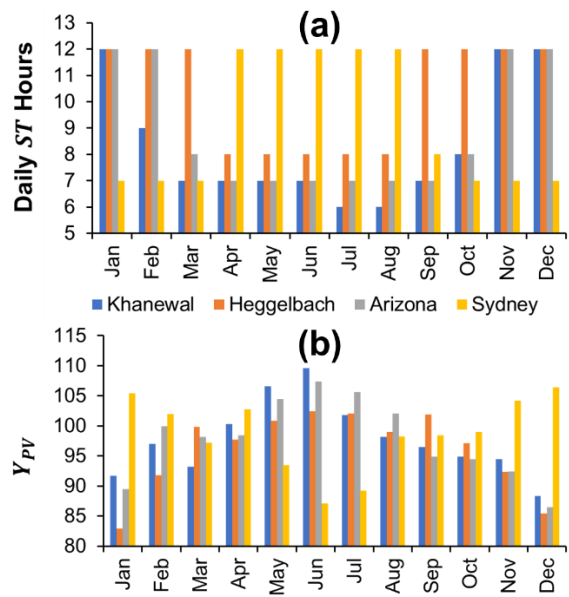

**Fig. A1.** Variation in monthly standard tracking hours and energy yield ($Y_{PV}$) for four different locations across the globe (Khanewal, Heggelbach, Arizona and Sydney). Energy yield is highest for summer months (June for northern hemisphere while Dec for southern hemisphere) even though the ST hours are less than 12 in these months.

B. *Crop sensitivities and revenue Inputs:*

In the high-value farm, crop rotation includes tomato, cauliflower, and garlic throughout the year, while the low-value farm involves wheat and cotton cultivation, as indicated in Table I. The economic details of these crops for 2018 in Pakistan are mentioned in Table I and are used for the economic case study in this paper. Fig. A2 shows the shade response for these crops as a function of land to area module ratio which is computed based on the model described in our previous study [36]. These crops can be categorized as shade-tolerant crops based on their $Y_{Crop}$ trend shown in Fig. A2 for $N/S$, $ST$ and $AT$ module systems.

Table I. Cropping cycle and net profit from Cotton and wheat for Low value farm, and Tomato, Cauliflower and Garlic for High Value Farm for Khanewal.

| Low value crops ($LV$) | | |
|---|---|---|
| **Months** | **Crop** | **Revenue ($/ha)[42]** |
| Apr-Sep | Cotton | 69.88 |
| Oct-Mar | Wheat | 228.43 |
| | Total | 298.31 |
| High value crops ($HV$) | | |
| **Months** | **Crop** | **Revenue ($/ha)[42]** |
| Apr-Jun | Tomato | 948.81 |
| Jul-Sep | Cauliflower | 1,145.98 |
| Oct-Mar | Garlic | 7,097.54 |
| | Total | 9,192.34 |

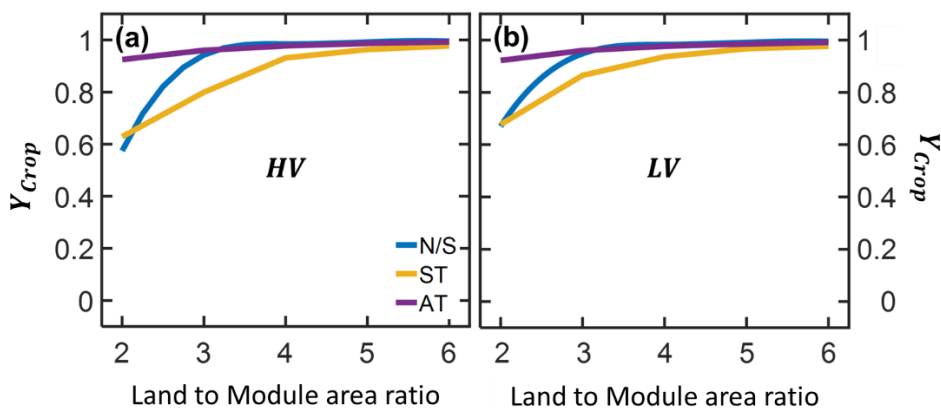

**Fig. A2.** Comparison of biomass yield ($Y_{Crop}$) of tomato, cauliflower, and ginger with shade tolerant crop for high value farm while cotton and wheat while shade tolerant crop for low value farm for N/S, ST and AT. These crops can be categorized as shade tolerant crops.

C. *Global variation in module to land soft cost ratio ($M_L$):*

Fig. A3. shows global variation in module to land soft cost ratio. $M_L$ is highest for Japan while lowest for Saudi Arabia and the average $M_L$ across the globe is around 10. Low values of $M_L$ corresponds to higher land costs [38]

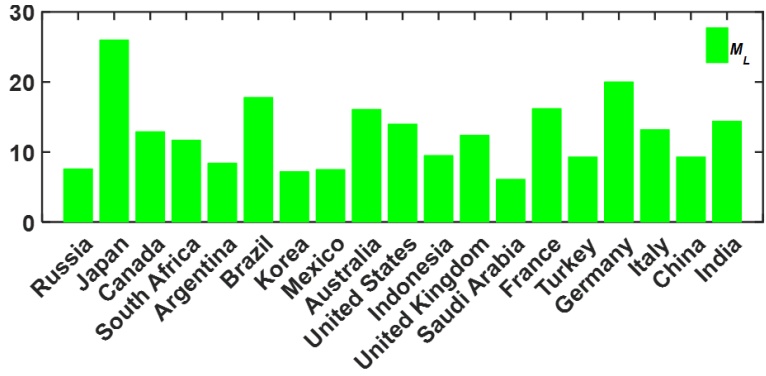

**Fig. A3.** Global variation in module to land soft cost ratio ($M_L$)[39]

## D. Effect of Feed-in tariff

Globally, electricity tariffs are on a downward trend, primarily driven by ongoing advancements in photovoltaic (PV) technology and the decreasing costs associated with it [43]. This phenomenon is also evident in Pakistan, where PV feed-in tariffs have been steadily declining, making solar energy more affordable [44]. In recent years, $PV$ tariffs in Pakistan have ranged from 5 to 7 cents per kilowatt-hour (KWh). To illustrate the impact of increasing the feed-in tariff ($FIT$) for agrivoltaics ($AV$) to cover the additional costs associated with AV systems, we present a table summarizing the $\Delta FIT$ (in %) required for $AV$ to achieve an economic equivalence with respect to $GMPV$ ($HV$ and $LV$ Farm) for Khanewal for different $A_{LM}$ and $M_L$ for $N/S$, $AT$ amd $ST$

Table II. ΔFIT (in %) required for AV to achieve an economic equivalence with respect to $GMPV$ ($HV$ and $LV$ Farm) for Khanewal

| $M_L$ | $A_{LM}$ | N/S LV | ST LV | AT LV | N/S HV | ST HV | AT HV |
|---|---|---|---|---|---|---|---|
| | | | | % Δ $FIT_{TH}$ | | | |
| 10 | 2 | 16.84 | 15.62 | 121.01 | 13.21 | 12.19 | 107.58 |
| | 4 | 21.41 | 19.59 | 131.60 | 8.92 | 9.39 | 103.23 |
| | 6 | 26.05 | 23.58 | 142.20 | 7.10 | 7.52 | 98.98 |
| 15 | 2 | 16.30 | 15.52 | 116.31 | 12.67 | 12.09 | 102.89 |
| | 4 | 19.26 | 18.09 | 123.21 | 6.76 | 7.89 | 94.84 |
| | 6 | 22.28 | 20.69 | 130.11 | 3.32 | 4.63 | 86.90 |
| 20 | 2 | 16.03 | 15.47 | 113.97 | 12.40 | 12.04 | 100.54 |
| | 4 | 18.18 | 17.35 | 119.01 | 5.68 | 7.15 | 90.64 |
| | 6 | 20.39 | 19.24 | 124.06 | 1.44 | 3.19 | 80.85 |
| 25 | 2 | 15.87 | 15.44 | 112.56 | 12.24 | 12.01 | 99.13 |
| | 4 | 17.53 | 16.90 | 116.50 | 5.03 | 6.70 | 88.12 |
| | 6 | 19.26 | 18.38 | 120.44 | 0.31 | 2.32 | 77.22 |
| 30 | 2 | 15.76 | 15.42 | 111.62 | 12.13 | 11.99 | 98.20 |
| | 4 | 17.10 | 16.60 | 114.82 | 4.60 | 6.40 | 86.44 |
| | 6 | 18.50 | 17.80 | 118.02 | 0.00 | 1.74 | 74.81 |

ACKNOWLEDGMENTS

This material is based upon work supported by the Doctoral Fellowship at LUMS. Authors also acknowledge the support of Professor Asharful Alam at Purdue University, USA for fruitful discussion and guidance throughout this work.